\documentclass[11pt]{article}

\usepackage{comment}
\usepackage{graphicx}
\usepackage[cp1250]{inputenc}
\usepackage[czech]{babel}
\usepackage{flafter}
\usepackage[nooneline,flushleft]{caption2}
\usepackage{color}
\usepackage{url}
\include{tdr-defs}

\begin{document}

\begin{center}
\noindent
{\LARGE Internal and external dynamics of antihydrogen in

\vskip 0.1cm
\noindent
electric and magnetic fields of arbitrary orientation}

\bigskip

Michal Špaček and Vojtěch Petráček

\textit{Department of Physics, Faculty of Nuclear Sciences and Physical Engineering, Czech Technical University in Prague, Břehová 7, 115 19 Prague 1, Czech Republic}

\bigskip

21.6.2012
\end{center}

\bigskip
\bigskip
\bigskip

We have studied the motion of antihydrogen atoms in electric and magnetic fields of arbitrary orientation. Our aim was to find an exact model of external (centre-of-charge) and internal dynamics which would allow us to describe antihydrogen motion in a realistic way. After analysing the kinematics of the charge exchange reaction producing antihydrogen and of antihydrogen deexcitation, we discuss the dynamics in cases of special fields configurations. Finally, we propose an expression for the force acting on an antihydrogen atom when the field configuration is arbitrary. The general formula -- when reduced in the special field configurations -- matches the results which we obtained independently before. The focus is both on the centre of charge dynamics and on the accurate internal quantum characteristics.

%We would like to model also antiatom manipulations such as antiatom acceleration, deceleration and trapping, antiatom beam focusation etc.

\section{Motivation and kinematics}

The motion of a charged particle in given arbitrary external electric and magnetic fields is governed by the Lorentz force as

$$ F_i = Q \left( E_i (\vec R) + \sum_{j, k = 1}^{3} \epsilon_{i j k} \dot{X}_j B_k (\vec R) \right) \mathrm{,} \eqno(1) $$

\noindent
where the $ i $, $ j $ and $ k $ indices label the Cartesian coordinates of the vectors. For a system of two coupled charges (like a hydrogen atom, an antihydrogen atom or a positronium), there is no such a reliable force in the given external fields yet, except for several special fields configurations. We have proposed such a force which is consistent with all crucial experimental and theoretical facts. Our model is simple enough to be solved analytically but yet sufficiently accurate (in the sense of the consistency) as well.

%As we focus on the dynamics of antihydrogen atoms in external electric and magnetic fields, we suppose the fields to be arbitrary oriented at any point of the space and time-independent (or at least varying very slowly when compared to the changes in internal characteristics of an atom) when treating the spectrum -- the processes on atomic scales are much faster than what the changes in external fields can cause. However, the whole atom (centre of charge) experiences, of course, the gradients of external fields. In these combined fields, we have tried to find the centre of charge dynamics of an individual antiatom. The particle is treated as point-like and its internal -- quantum -- properties (not necessarily independent of the external fields) are taken into consideration all the time.

As we focus on the dynamics of antihydrogen atoms in external electric and magnetic fields, we suppose the fields to be arbitrary oriented at any point of the space and time-independent (or at least varying very slowly when compared to the changes in internal characteristics of an atom). When we deal with the atom spectrum, we consider external fields to be completely constant in time (even when the atom is in motion) since the speed of inner atomic processes makes changes in external fields negligible; when studying atom as a whole, we consider all possible spatial derivatives of external fields. In these combined fields, we have tried to find the centre of charge dynamics of an individual antiatom. The particle is treated as point-like and its internal -- quantum -- properties (not necessarily independent of the external fields) are taken into consideration all the time.

From the currently available technologies we consider the charge exchange reaction to be the best method of the cold antihydrogen production. In the charge exchange, an antiproton replaces an electron in the positro-nium in a highly excited Rydberg state. As a product, one obtains an excited Rydberg antihydrogen atom:

$$ \mathrm{Ps} (n_{\mathrm{Ps}}) + p^{-} \rightarrow \mathrm{\bar{H}} (n) + e^{-} \eqno(2) $$

\noindent
The principal quantum numbers of a positronium and an antihydrogen are denoted as $ n_{\mathrm{Ps}} $ and $ n $ respectively. The kinematics of the reaction provides us with the initial conditions of the antihydrogen atom motion determining the magnitude and space orientation of the velocity vector $ v_{\bar{\mathrm{H}}} $ immediately after the charge exchange event. The kinematics of the charge exchange process is described by the equation [1]

$$ v_{\mathrm{\bar{H}}}^2 - 2 \left ( v_{\bar{p}} \cos \theta_{\mathrm{\bar{H}}} + 2 \frac{m_e}{m_{\bar{p}}} v_{\mathrm{Ps}} \cos ( \theta_{\mathrm{Ps}} – \theta_{\mathrm{\bar{H}}} ) \right ) v_{\mathrm{\bar{H}}} + $$

$$ + 2 \left ( \frac{m_e^2}{m_{\bar{p}}^2} v_{\mathrm{Ps}}^2 + 2 \frac{m_e}{m_{\bar{p}}} v_{\bar{p}} v_{\mathrm{Ps}} \cos \theta_{\mathrm{Ps}} – \frac{m_e}{m_{\bar{p}}^2} R_{\mathrm{\bar{H}}} \left ( \frac{1}{n^2} - \frac{1}{2 n_{\mathrm{Ps}}^2} \right ) \right ) = 0 \mathrm{,} \eqno(3) $$

\noindent
where $ m_X $ and $ v_X $ stand for masses and velocities of the particles respectively and $ \theta_{X} = \angle \left( \vec{v}_{X}, \vec{v}_{\mathrm{\bar p}} \right) $; $ R_{\bar H} $ denotes the Rydberg constant.

By analysing (3) as a hyperplane of all possible states $ (v_{\mathrm{\bar H}}, \theta_{\mathrm{\bar H}}) $, we found that the angle $ \theta_{\mathrm{\bar H}} $ does not significantly deviate from zero in any kinematics configuration that means the direction of its motion only slightly differs from original direction of the antiproton motion.

\newpage

What remains, as far as the kinematics is concerned, is to consider the effect of the antihydrogen deexcitation, that is

$$ \bar{H} (n) \rightarrow \bar{H} (n') + \gamma \mathrm{.} \eqno(4) $$

\noindent
We found the change in the antihydrogen velocity magnitude to be

$$ \Delta v' \approx \frac{R}{m_{\bar p} c^2 + m_{\bar e} c^2} \left ( \frac{1}{{n'}^2} - \frac{1}{n^2} \right) c \mathrm{.} \eqno(5) $$

When an antihydrogen deexcitation event occurs, the change in the velocity magnitude of the atom is negligible. However the change of $ n $ into $ n' $ itself has to be taken into consideration since the dynamics in external fields depends on the internal state of the antihydrogen.

\section{Antihydrogen dynamics}

In our model, we assume that the dipole qualities of an antihydrogen atom are decisive for its macroscopic dynamics. A hydrogen atom -- and therefore an antihydrogen atom, too -- evinces four analytically different kinds of dipoles: a not-induced electric dipole, a not-induced magnetic dipole, an induced electric dipole and an induced magnetic dipole. Moreover, one has to distinguish a not-induced orbital magnetic dipole and a not-induced spin magnetic dipole. When speaking about spin, we always refer to the spin of the positron; the spin of the antiproton is neglected in our model.

We claim that a classical not-induced electric dipole $ \vec d $ (for instance a water molecule) placed in a stationary external electric field $ \vec E = \vec E ( \vec R ) $ experiences a force expressed in the Cartesian coordinates as

$$ F_i = \tilde d \sum_{j = 1}^{3} \frac{E_j}{E} \frac{\partial E_j}{\partial X_i} \mathrm{,} \eqno(6) $$

\noindent
where $ \tilde d $ stands for the projection of the dipole vector $ \vec d $ into the direction given by the local vector $ \vec E $. When the quantum nature of an antihydrogen atom is to be effectively implemented, the $ \tilde d $ quantity is the function of the proper quantum numbers.

The force acting on a dipole can be derived using two approaches. The first method treats the dipole as an object of a finite size -- this corresponds to the separation of two charges defining the dipole. The system has more than three degrees of freedom which may be however reduced: The torque equation $ \vec N = \vec d \times \vec E $ prescribes that $ \vec E $ is the local preferred direction around which the $ \vec d $ vector undergoes internal motion with a well-defined time-averaged value $ \tilde d $. This holds for time-independent external fields. The force on this particularly defined dipole is therefore fully responsible for the dipole's macroscopic motion and equals (6). This approach was inspired by outcomes in [2] and by unpublished result [3].

The second possibility is to take the classical expression $ W = - \tilde d E $ as the potential energy of a not-induced electric dipole in an external electric field (with the quantum conditions still easily applicable on $ \tilde d $) [4]. In this case, the force is

$$ F_{i} = - \frac{\partial}{\partial X_i} \left( - \tilde d E \right) = \tilde d \frac{\partial}{\partial X_i} E = \tilde d \frac{\partial}{\partial X_i} \sqrt{ \sum_{j = 1}^3 E_j E_j } = \tilde d \frac{2 \sum_{j = 1}^3 E_j \frac{\partial E_j}{\partial X_i}}{2 \sqrt{ \sum_{j = 1}^3 E_j E_j }} = $$

$$ = \frac{\tilde d}{\sqrt{ \sum_{j = 1}^3 E_j E_j }} \sum_{j = 1}^3 E_j \frac{\partial E_j}{\partial X_i} = \tilde d \sum_{j = 1}^3 \frac{E_j}{E} \frac{\partial E_j}{\partial X_i} \mathrm{.} \eqno(7) $$

The situation for a not-induced magnetic dipole $ \vec \mu $ placed in an external magnetic field $ \vec B = \vec B ( \vec R ) $ is similar: Its potential energy $ W = - \tilde \mu B $ leads to the same force on the dipole

$$ F_i = \tilde \mu \sum_{j = 1}^{3} \frac{B_j}{B} \frac{\partial B_j}{\partial X_i} \eqno(8) $$

\noindent
as it is obtained by solving the force on an elementary current loop. The $ \tilde \mu $ stands for the projection of $ \vec \mu $ into that preferred $ \vec B $ direction. Moreover, the internal dynamics is analytically solvable and known as the Larmor precession.

For induced electric and magnetic dipoles separately, the dynamics is very similar: The forces resemble (6) and (8), only the dipoles are explicitly proportional to the field magnitude with accordance to their definition. In these cases we use $ \alpha E $ instead of $ \tilde d $ in (6) and $ \beta B $ instead of $ \tilde \mu $ in (8) respectively.

The result (6) is also applicable to the antihydrogen atom in such a field if $ \tilde d $ is substituted with a proper function of the internal quantum numbers of the antihydrogen. The shifts of spectral lines of the antihydrogen atom in an electric field [5] equal

$$ W = \frac{3}{2} a_0 e n p E \mathrm{,} \eqno(9) $$

\noindent
where $ a_0 $ denotes the Bohr radius, $ e $ is the elementary charge, $ n $ is the principal quantum number of the atom and $ p $ stands for the parabolic quantum number (an element of the $ \{ - n + 1, - n + 2, ..., n - 2, n - 1 \} $ set). If compared to $ W = - \tilde d E $, one has $ \tilde d = - \frac{3}{2} a_0 e n p $ and the force on such a particular anti-

\noindent
atom in an external electric field is

$$ F_i = - \frac{3}{2} a_0 e n p \sum_{j = 1}^{3} \frac{E_j}{E} \frac{\partial E_j}{\partial X_i} \mathrm{,} \eqno(10) $$

\noindent
which is (6) in fact, with the quantum character of the atom implemented.

The other three dipole cases for the particular antiatom can be evaluated in very similar way. Namely for that of a not-induced magnetic dipole in a magnetic field, the energy (Zeeman effect) shifts [6] are

$$ W = \frac{e \hbar}{2 m_e} m B \mathrm{,} \eqno(11) $$

\noindent
where $ m_e $ is the rest mass of an electron, $ \hbar $ is the reduced Planck constant and $ m $ stands for the magnetic quantum number (for a given $ n $ it is an element of the same set as the parabolic quantum number is). When compared to $ W = - \tilde \mu B $, one obtains

$$ F_i = - \frac{e \hbar}{2 m_e} m \sum_{j = 1}^{3} \frac{B_j}{B} \frac{\partial B_j}{\partial X_i} \mathrm{,} \eqno(12) $$

\noindent
when $ \tilde \mu = - \frac{e \hbar}{2 m_e} m $ is substituted into (8). Note that both (11) and (12) hold only if the spin of the electron (positron) is neglected -- the effect of spin will be discussed later.

The antihydrogen spectra themselves prove to be reliable starting point for treating the atom dynamics. The question now is, how these energy shifts look like when both electric and magnetic fields are present (that is which expression reduces to the (9) and (11) energies in the single-field limits). The expression for $ W $ should be explicitly dependent on the angle $ \gamma $ between the two local field orientations ($ \gamma = \angle (\vec E, \vec B) $) [7].

The most straightforward way to construct the antihydrogen spectra would be a simple sum of (9) and (11) -- the Stark and Zeeman energies. However, such an object, described by

$$ W = - \frac{3}{2} a_0 e n p E - \frac{e \hbar}{2 m_e} m B \mathrm{,} \eqno(13) $$

\noindent
does not satisfy the condition of the angular dependence and is therefore incorrect. The accurate expression has to be found -- this is the core of this paper.

With the spin still not taken into account, for any local electromagnetic field, such a Cartesian coordinate system can be found that the most basic Hamiltonian [8] of a hydrogen atom in the simultaneous electric and magnetic fields reads

$$ \hat H = \frac{\hat{\vec p}^{~2}}{2 m_e} - \frac{e^2}{4 \pi \epsilon_0 \hat r} + \frac{e \hbar}{2 m_e} B \hat{L}_3 - e E \left( \hat{x} \sin \gamma + \hat{z} \cos \gamma \right) \mathrm{.} \eqno(14) $$

\noindent
The perturbation theory leads to (9) when $ B = 0 $ is set in (14) and we get (11) for $ E = 0 $ as well. Unfortunately, there is no known way for a general fields configuration how to analytically evaluate the spectrum of the matrix whose elements are given by the integral

%Unfortunately, the whole (14) Hamiltonian is not analytically solvable within the perturbation theory since we are not able to evaluate the spectrum of the matrix whose elements are given by the integral

$$ W_1^{(l' m')(l m)} = \int R_{n l'}^{*} Y_{l' m'}^{*} \left( + \frac{e}{2 m_e} B \hat{L}_{3} - e E \hat{x}_3 \right) R_{n l} Y_{l m} ~\mathrm{d} V \mathrm{,} \eqno(15) $$

\noindent
where $ \left( \frac{\hat {\vec p}^{~2}}{2 m_{e}} - \frac{e^2}{4 \pi \epsilon_0 \hat r} \right) \left( R_{n l} Y_{l m} \right) = - \frac{R}{n^2} \left( R_{n l} Y_{l m} \right) $.

There are alternative methods of finding the spectra, namely the Sommerfeld quantum conditions which give exactly the same results as (9) and (11). What makes them important is that the combined problem has been already solved using them, as described by Max Born in [9], for instance. The final spectrum reads

$$ W_1 = \left| + \frac{3 a_0 e n}{2 \hbar} \vec E + \frac{e}{2 m_e} \vec B \right| n_A \hbar + \left| - \frac{3 a_0 e n}{2 \hbar} \vec E + \frac{e}{2 m_e} \vec B \right| n_B \hbar \mathrm{,} \eqno(16) $$

\noindent
where two new quantum numbers $ n_A $ and $ n_B $ (instead of $ p $ and $ m $) were introduced. Both $ n_A $ and $ n_B $ are independently elements of the set \\$ \left \{ - \frac{n - 1}{2}, - \frac{n - 1}{2} + 1, ..., + \frac{n - 1}{2} - 1, + \frac{n - 1}{2} \right \} $.

Due to the fact that (16) does not reflect the effect of the spin of the electron, the outcome has to be slightly modified. It seems impossible to implement the effect generally, even in the absence of the electric field. There are in fact three distinct not-induced dipoles in a hydrogen atom: not-induced electric dipole, not-induced orbital magnetic dipole and not-induced spin magnetic moment which have to be distinguished. The expression (16) connects the first two together with no limitation on the field strengths. On the contrary, the connection of the two not-induced magnetic dipoles depends on the field strength [8]. However, to maintain the ambition of a universal equation of motion, we suggest the spin effect in a way we consider simple and minimally harming the accuracy -- by adding the term $ \frac{e \hbar}{m_e} s B $ (where $ s = \pm \frac{1}{2} $) to (16). Moreover, when the absolute values are expanded, also the explicit angular dependence on $ \gamma $ becomes

$$ W = \sqrt{\left ( \frac{3}{2} a_0 e n E \right )^2 + \left ( \frac{e \hbar}{2 m_e} B \right )^2 + \frac{3 a_0 e^2 n \hbar}{2 m_e} E B \cos \gamma} \cdot n_A + $$

$$ + \sqrt{\left ( \frac{3}{2} a_0 e n E \right )^2 + \left ( \frac{e \hbar}{2 m_e} B \right )^2 - \frac{3 a_0 e^2 n \hbar}{2 m_e} E B \cos \gamma} \cdot n_B + \frac{e \hbar}{m_e} s B \mathrm{.} \eqno(17) $$

%The reason why it is not possible to evaluate (15) analytically may be connected with the fact that the two particular perturbations $ + \frac{e}{2 m_e} B \hat{L}_3 $ and $ - e E \hat{x}_3 $ do not commute [10].

We found that

$$ \left[ \frac{e}{2 m_e} \vec B \hat{\vec L}, e \vec E \hat{\vec r} \right] = - i \hbar \frac{e^2}{2 m_e} \left( \vec E \times \vec B \right) \vec r \mathrm{.} \eqno(18) $$

\noindent
This relation also implies that the associated quantum numbers $ p $ and $ m $ are well-defined only when commutator (18) equals zero, that is in the pure Stark and Zeeman cases, for instance [10]. Therefore the new quantum numbers $ n_A $ and $ n_B $ were introduced instead of $ p $ and $ m $ (in our model, the spin quantum number $ s $ is well-defined for any fields configuration). The total spin-less degeneracy of the $ n $-th manifold equals

$$ \sum_{l = 0}^{n - 1} \sum_{m = - l}^{+l} 1 = \sum_{l = 0}^{n - 1} \left ( 2 l + 1 \right ) = 2 \sum_{l = 0}^{n - 1} l + \sum_{l = 0}^{n - 1} 1 = 2 \cdot \frac{ n \left ( n - 1 \right ) }{2} + n = n^2 \mathrm{,} \eqno(19) $$

\noindent
which has to be satisfied for the new quantum numbers as well and

$$ \sum_{n_A = - \frac{n - 1}{2}}^{+ \frac{n - 1}{2}}~ \sum_{n_B = - \frac{n - 1}{2}}^{+ \frac{n - 1}{2}} 1 = \left( \sum_{n_A = - \frac{n - 1}{2}}^{+ \frac{n - 1}{2}} 1 \right) \cdot \left( \sum_{n_B = - \frac{n - 1}{2}}^{+ \frac{n - 1}{2}} 1 \right) = n \cdot n = n^2 \mathrm{.} \eqno(20) $$

%\noindent
%This demonstrates the good quality of the new quantum numbers.

The Stark and Zeeman forces are two special cases of the (16) or (17) results. Note that instead of (12), where the effect of the spin was not taken

\newpage
\noindent
into account, we obtain

$$ F_i = - \frac{e \hbar}{2 m_e} \left( m + 2 s \right) \sum_{j = 1}^{3} \frac{B_j}{B} \frac{\partial B_j}{\partial X_i} \mathrm{.} \eqno(21) $$

\noindent
Aside of the Stark and Zeeman effects, we found two more special cases. The first is the case when the electric and magnetic fields are parallel -- according to (18), the two perturbations are independent and the "old" quantum numbers $ p $ and $ m $ are well-defined. The potential energy in this case is therefore the simple sum (13) modified by the term $ \frac{e \hbar}{m_e} s B $ responsible for the spin effect. Hence the force on an atom in parallel fields is the negative gradient of that potential energy and reads as

$$ F_i = - \frac{e \hbar}{2 m_e} \left( m + 2 s \right) \sum_{j = 1}^{3} \frac{B_j}{B} \frac{\partial B_j}{\partial X_i} - \frac{3}{2} a_0 e n p \sum_{j = 1}^{3} \frac{E_j}{E} \frac{\partial E_j}{\partial X_i} \mathrm{.} \eqno(22) $$

The final case which we were able to solve independently of (17) is such configuration of fields when the magnetic field is much stronger than the electric field. We analysed this case using the fact that the magnetic field is dominant and therefore we constructed the potential energy as $ W = \frac{e \hbar}{2 m_e} \left( m + 2 s \right) B + \frac{3}{2} a_0 e n p E \cos \gamma $ -- the angle dependence is present at the electric term since the orientation of the $ \vec B $ vector is significant for the atom and the second term as whole is considered to be a small correction. The force on an atom in such fields is

%The final case which we were able to solve independently of (17) is such configuration of fields when the magnetic field is much stronger than the electric field. We analysed this case by a double-step perturbation theory and found the potential energy to be $ W = \frac{e \hbar}{2 m_e} \left( m + 2 s \right) B + \frac{3}{2} a_0 e n p E \cos \gamma $ -- the angle dependence is present at the electric term since the magnetic one is dominant and the orientation of the $ \vec B $ vector is significant for the atom. The force on an atom in such fields is

$$ F_i = - \frac{e \hbar}{2 m_e} \left( m + 2 s \right) \sum_{j = 1}^{3} \frac{B_j}{B} \frac{\partial B_j}{\partial X_i} - \frac{3}{2} a_0 e n p \sum_{j = 1}^{3} \frac{E_j}{E} \frac{\partial E_j}{\partial X_i} \cos \gamma \mathrm{.} \eqno(23) $$

Now let us return to the general case for which the potential energy is given by (17). In very strong fields, the effects of the induced dipoles (quadratic effects) have to be taken into account. We assume that the two terms $ - \frac{1}{2} \alpha E^2 $ and $ - \frac{1}{2} \beta B^2 $ responsible for them can be simply added to the potential energy since their effects are independent of the not-induced dipoles in the first approximation. The polarisabilities $ \alpha $ and $ \beta $ depend in a bit more complicated way on $ n_A $, $ n_B $ and $ n $. The total external force (as the negative gradient of the potential energy) on such a model of an atom therefore is:

\newpage

$$ F_i = - \frac{\left ( \frac{3}{2} a_0 e n \right )^2 \sum_{j = 1}^{3} E_j \frac{\partial E_j}{\partial X_i} + \left ( \frac{e \hbar}{2 m_e} \right )^2 \sum_{j = 1}^{3} B_j \frac{\partial B_j}{\partial X_i}}{\sqrt{\left ( \frac{3}{2} a_0 e n E \right )^2 + \left ( \frac{e \hbar}{2 m_e} B \right )^2 + \frac{3 a_0 e^2 n \hbar}{2 m_e} E B \cos \gamma}} n_A $$

$$ - \frac{\frac{3 a_0 e^2 n \hbar}{4 m_e} \cos \gamma \sum_{j = 1}^{3} \left( B \frac{E_j}{E} \frac{\partial E_j}{\partial X_i} + E \frac{B_j}{B} \frac{\partial B_j}{\partial X_i} \right)}{\sqrt{\left ( \frac{3}{2} a_0 e n E \right )^2 + \left ( \frac{e \hbar}{2 m_e} B \right )^2 + \frac{3 a_0 e^2 n \hbar}{2 m_e} E B \cos \gamma}} n_A $$

$$ - \frac{\left ( \frac{3}{2} a_0 e n \right )^2 \sum_{j = 1}^{3} E_j \frac{\partial E_j}{\partial X_i} + \left ( \frac{e \hbar}{2 m_e} \right )^2 \sum_{j = 1}^{3} B_j \frac{\partial B_j}{\partial X_i}}{\sqrt{\left ( \frac{3}{2} a_0 e n E \right )^2 + \left ( \frac{e \hbar}{2 m_e} B \right )^2 - \frac{3 a_0 e^2 n \hbar}{2 m_e} E B \cos \gamma}} n_B $$

$$ + \frac{\frac{3 a_0 e^2 n \hbar}{4 m_e} \cos \gamma \sum_{j = 1}^{3} \left( B \frac{E_j}{E} \frac{\partial E_j}{\partial X_i} + E \frac{B_j}{B} \frac{\partial B_j}{\partial X_i} \right)}{\sqrt{\left ( \frac{3}{2} a_0 e n E \right )^2 + \left ( \frac{e \hbar}{2 m_e} B \right )^2 - \frac{3 a_0 e^2 n \hbar}{2 m_e} E B \cos \gamma}} n_B $$

$$ - \frac{e \hbar}{m_e} s \sum_{j = 1}^{3} \frac{B_j}{B} \frac{\partial B_j}{\partial X_i} + \alpha \sum_{j = 1}^{3} E_j \frac{\partial E_j}{\partial X_i} + \beta \sum_{j = 1}^{3} B_j \frac{\partial B_j}{\partial X_i} \mathrm{.} \eqno(24) $$

What makes us confident about this result is that all the independently solvable cases -- the outcomes (10), (21), (22) and (23) -- are exactly the limits of (24) for the particular fields configurations. In the four independently solvable cases, we found that (and how) the quantum numbers $ p $ and $ m $ can be linked to $ n_A $ and $ n_B $.

All the terms are proportional to the field derivatives hence there is no force on an atom in homogeneous fields. All the weak field terms are                                                                                                                                         antisymmetric in $ n_A $, $ n_B $ and $ s $. This allows using the formula (24) for both matter and antimatter with only the redefinition of these quantum numbers.

\section{Conclusion}

We have analysed the spectrum of an antihydrogen atom in time-independent external electric and magnetic fields of arbitrary mutual orientation. We have discussed the kinematics of a possible deexcitation and the kinematics of a charge exchange reaction -- which we consider the best method for the cold antihydrogen production. Charge exchange kinematics determines the initial conditions for the motion of the antihydrogen.

In the search for the antihydrogen dynamics itself, we studied the ele-mentary and analytically solvable cases, such as an electric dipole in an external electric field for which the Stark-like hydrogen is a special case. In this sense we have justified the spectral approach to be reliable and equi-valent to the potential energy since the antihydrogen possesses three degrees of freedom in our model. We used the result derived by Max Born [9] to be such an accurate spectral term for an antihydrogen atom and we implemented the effect of a spin and possible strong-field effects. The potential energy satisfies all the experimental properties and so does the external force acting on the antiatom, which is (24).

We hope that the equation will prove itself useful and it will help to design experimental devices and carry out measurements on antihydrogen atoms. We are especially interested in the gravitational properties of antimatter and we hope that the theoretical understanding of the antiatom motion in the combined fields will contribute to the precise understanding of systematic errors in measurements of the atihydrogen in free fall after its acceleration. If we succeed in effective trapping of antiatoms, much more experiments would be possible. The same holds for positronium.

\section*{Acknowledgment}

For their interest, advice and comments which helped to enhance the quality of the paper, we would like to thank to our colleagues prof. Ing. Igor Jex, DrSc., RNDr. Miloš Pachr, CSc., RNDr. David Břeň, Ph.D. and Ing. Václav Potoček.

\section*{References}

\noindent
[1] ŠPAČEK, Michal. \textit{Dynamics of anti-hydrogen motion in the AEGIS experiment}. Prague, 2012. Diploma thesis. Czech Technical University in Prague, Faculty of Nuclear Sciences and Physical Engineering, Department of Physics. Supervised by doc. RNDr. Vojtěch Petráček, CSc.

\noindent
[2] FORMÁNEK, Jiří. \textit{Úvod do kvantové teorie}. First edition. Prague: Academia, 1983. ISBN 80-200-1176-5.

\noindent
[3] \textit{Force and Torque on a Small Magnetic Dipole}. Physics Insights [online]. [cit. 2012-01-04]. Available from: \url{http://www.physicsinsights.org/force_on_dipole_1.html}.

\noindent
[4] SEDLÁK, Bedřich; ŠTOLL, Ivan. \textit{Elektřina a magnetismus}. Prague : Academia, 2002. 636 p. ISBN 80-200-1004-1.

\newpage

\noindent
[5] GILMORE, Robert. \textit{Stark Effect}. Drexel University [online]. [cit. 2012-01-04]. Available from: \url{http://www.physics.drexel.edu/~bob/PHYS517_10/stark.pdf}.

\noindent
[6] LANDAU, Lev Davidovich; LIFSHITZ, Evgeny Mikhailovich. \textit{Quantum mechanics : Non-relativistic theory}. Third edition, revised and enlarged. [Oxford] : Pergamon Press, 1977. 677 p. ISBN 0-08-020940-8.

\noindent
[7] Private communication: Dr. Stephen Hogan, ETH Zürich

\noindent
[8] COHEN-TANNOUDJI, Claude, DIU, Bernard, a LALOË, Frank. \textit{Quantum mechanics}. Paris: Collection Enseignement des sciences, 1973. ISBN 0-471-16433-X.

\noindent
[9] BORN, Max. \textit{Vorlesungen über Atommechanik}. Berlin: Springer, 1925.

\noindent
[10] Private communication: Ing. Václav Potoček, CTU Prague

\end{document}